\documentclass[fleqn,12pt,twoside]{article}
\usepackage[headings]{espcrc1}

\readRCS $Id: espcrc1.tex,v 1.2 2004/02/24 11:22:11 spepping Exp $
\usepackage{graphicx}
\usepackage{epsfig}
\usepackage{amssymb}
\usepackage[figuresright]{rotating}

\newcommand{\AmS}{{\protect\the\textfont2
  A\kern-.1667em\lower.5ex\hbox{M}\kern-.125emS}}

\title{{Complexiton solutions to integrable equations}}

\author{Wen-Xiu Ma\address[MCSD]{Department of Mathematics,
University of South Florida, Tampa, FL 33620-5700, USA
}%
        \thanks{Email address: mawx@math.usf.edu. This work was supported in part by the University of South Florida
Internal Awards Program under Grant No. 1249-936RO.}
                }

\runtitle{Complexiton Solutions}
 \runauthor{W.X. Ma}

\begin{document}

\maketitle

\def \bea {\begin{eqnarray}}
\def \eea {\end{eqnarray}}
\def \ba {\begin{array}}
\def \ea {\end{array}}
\def \part {\partial}
\def \D {\displaystyle }
\newcommand{\R}{\mathbb{R}}

\begin{abstract}
Complexiton solutions (or complexitons for short) are exact
solutions newly introduced to integrable equations. Starting with
the solution classification for a linear differential equation,
the Korteweg-de Vries equation and the Toda lattice equation are
considered as examples to exhibit complexiton structures of
nonlinear integrable equations. The crucial step in the solution
process is to apply the Wronskian and Casoratian techniques for
Hirota's bilinear equations. Correspondence between complexitons
of the Korteweg-de Vries equation and complexitons of the Toda
lattice equation is provided.
\end{abstract}

\section{INTRODUCTION}

Differential equations or differential-difference equations can
describe various motions in nature. It is important to study their
integrable properties, and more important to tangibly determine
their exact solutions. Soliton theory is one of significant
developments along this direction. The theory tells us that there
exist soliton solutions to many integrable equations, both
continuous and discrete. More generally, negatons (generalized
solitons) can be explicitly presented (see
\cite{RasinariuSK-JPA1996,MarunoMO-JPSJ2004}, for example, for the
Korteweg-de Vries (KdV) equation and the Toda lattice equation).
Similarly, there exist positons
\cite{Matveev-PLA1992}-\cite{Ma-CSF2004}, which is another
important achievement in soliton theory. What could we say about
exact solutions further? This report is aiming at discussing this
question. Specifically, we will explore what other solutions to
integrable equations can exist, and show that so-called
complexiton solutions \cite{Ma-PLA2002} are one of new exact
solutions.

Let us first observe an example of linear ordinary differential
equations:
\begin{equation}
 ay''+by'+cy=0,\quad y'=dy/dx,\quad y''=d^2y/dx^2,\quad a,b,c=\textrm{real consts.},
\label{eq:2ndorderLDE:pma-wcna2004}
\end{equation}
to recall the solution classification of linear differential
equations.
 Its characteristic
equation is a quadratic equation
\[ am ^2+bm +c=0,\] and the quadratic formula gives
 {its two roots:}
 \[m _{1,2}=\frac {-b\pm \sqrt {b^2-4ac}}{2a}.\]
These two roots completely determine the general solution $
y=y(x)$ of (\ref{eq:2ndorderLDE:pma-wcna2004}). We now describe
all solution situations of (\ref{eq:2ndorderLDE:pma-wcna2004}),
with pointing out the corresponding solutions in soliton theory.
\begin{itemize}
\item {\bf Real roots:}
\begin{itemize}
\item {Distinct roots} $m _{1}\ne m _2$: The general solution of
(\ref{eq:2ndorderLDE:pma-wcna2004}) is then
\[y(x)=c_1\textrm{e}^{m  _1x}+c_2 \textrm{e}^{m  _2x},\quad c_1,c_2\in \R .\]
This corresponds to so-called solitons and negatons in soliton
theory.
 \item {Repeated roots} $m _{1}= m _2$: The general solution of
(\ref{eq:2ndorderLDE:pma-wcna2004}) is now
\[y(x)=c_1\textrm{e}^{m  _1x}+c_2 x\textrm{e}^{m  _1x},\quad c_1,c_2\in \R .\]
This corresponds to negatons of higher order when $m_1=m_2\ne 0$
and rational solutions when $m_1=m_2=0$ in soliton theory.
\end{itemize}

\item {\bf Complex roots:}
\begin{itemize}
\item {Purely imaginary roots} $m _{1,2}=\pm \beta \sqrt{-1}$: The
general solution of (\ref{eq:2ndorderLDE:pma-wcna2004}) is then
\[y(x)=c_1\sin
(\beta x)+c_2 \cos (\beta x),\quad c_1,c_2\in \R .\] {This
corresponds to so-called positons in soliton theory}.

 \item { Not purely imaginary roots} $m _{1,2}=\alpha
\pm \beta \sqrt{-1}$: The general solution of
(\ref{eq:2ndorderLDE:pma-wcna2004}) is now
  \[y(x)=c_1\textrm{e}^{\alpha x}\sin
(\beta x)+c_2 \textrm{e}^{\alpha x}\cos (\beta x),\quad c_1,c_2\in
\R .\] { This corresponds to so-called complexitons which we are
going to discuss.} This solution can also boil down to periodic
solutions of positon type if $\alpha \to 0$, and exponential
function solutions of negaton type if $\beta \to 0$.

\end{itemize}

\end{itemize}

It is always possible to classify exact solutions of
constant-coefficient, linear ordinary differential equations of
any order. A great review of solution classifications was given in
Ince's book \cite{Ince-book1956}, where the solutions of the
second-order linear ordinary differential equations were
classified in terms of hypergeometric functions, Riemann
P-functions, etc. The problem of solution classifications becomes
very difficult for nonlinear differential equations. Galois
differential theory may help in handling differential equations of
polynomial type. We will only concentrate on integrable equations,
for which we can start from their nice mathematical properties
such as symmetries and adjoint symmetries
\cite{MaS-PLA1994,MaZ-JNMP2002}.

{\bf What is a complexiton solution?} The general notion of
complexitons could be characterized by the following two criteria:
\begin{itemize}
\item Complexitons involve two kinds of transcendental functions:
 trigonometric
 and exponential functions.
\item Complexitons correspond to complex eigenvalues of associated
characteristic problems.
\end{itemize}
This report aims at constructively contributing to the theory of
complexiton solutions to nonlinear integrable equations, and the
KdV equation and the Toda lattice equation will be taken as two
illustrative examples. The proposed idea of constructing
complexitons through special determinants, for example, the
Wronskian and Casorati determinants, will also work for other
integrable equations. In particular, super-complexitons can be
similarly generated for super integrable equations, which will
extend the theory of supersolitons
\cite{IbortMM-JMP1996,LiuM-PLB1998}.

\section{KORTEWEG-DE VRIES EQUATION}

Let us first consider the KdV equation
\begin{equation}
 u_{t}+u_{xxx}-6uu_x=0,\label{eq:kdv:pma-wcna2004}
\end{equation}
where $u_t=\frac {\partial u}{\partial t}, \, u_x=\frac {\partial
u}{\partial x}$ and $u_{xxx}=\frac {\partial ^3u}{\partial x^3}$.
It is known that
  under
 the transformation
\begin{equation} u=- 2(\textrm{ln}f)_{xx}=-\frac {2(ff_{xx}-f_x^2)}{f^2},
\end{equation}
the KdV equation (\ref{eq:kdv:pma-wcna2004}) is transformed into
the bilinear equation
\[  (D_xD_t+D_x^4)f\cdot
f=0,\] that is,
\begin{equation}
 f_{xt}f-f_tf_x+f_{xxxx}f-4f_{xxx}f_x+3f_{xx}^2=0, \label{eq:bkdv:pma-wcna2004}
 \end{equation}
where $D_x$ and $D_t$ are Hirota's operators:
\[ f(x+h,t+k)g(x-h,t-k)
=\sum_{i,j=0}^\infty \frac 1 {i!j!}(D_x^iD_t^jf\cdot g)h^ik^j,
\]
 or more directly,
  \[ D_x^mD_t^n(f\cdot g)=
\left(\frac  \partial {\partial x}- \frac \partial {\partial
x'}\right)^m \left(\frac  \partial {\partial t}- \frac \partial
{\partial t'}\right)^n f(x,t)g(x',t')\Bigl.\Bigr|_{x'=x,t'=t}.\]

A powerful method of solutions for integrable bilinear equations
is the Wronskian technique \cite{Satsuma-JPSJ1979}. To construct
solutions, we use the Wronskian determinant
\begin{equation}
 f=W(\phi _1,\phi_2,\cdots,\phi_{N}):=\left |
\ba {cccc}\phi _1^{(0)}&\phi _1^{(1)}&\cdots &\phi_1^{(N-1)}\\
\phi _2^{(0)}&\phi_2^{(1)}&\cdots
 &\phi_2^{(N-1)}\\ \vdots &\vdots & \ddots & \vdots \\
\phi _{N}^{(0)}&\phi_{N}^{(1)}&\cdots
 &\phi_N^{(N-1)}\ea
\right |,\label{eq:wronskian:pma-wcna2004}\end{equation}
 where
$ \phi _i^{(j)}= {\part ^j\phi_i}/{\part x^j},\ j\ge 0.$ The
resulting solutions are called Wronskian solutions. The Wronskian
technique requires
 \[ -\phi_{i,xx}=\lambda _i \phi_i,\quad \phi_{i,t}=-4\phi_{i,xxx},\quad 1\le i\le N, \]
 and thus all involved functions $\phi_i$, $1\le i\le N$, are
 eigenfunctions of the Lax pair of the KdV equation
 associated with
zero potential. Actually, the Wronskian solution can be generated
from the Darboux transformation of the KdV equation starting with
zero solution. The above system generates the eigenfunctions
needed in forming Wronskian solutions:
\bea  \phi_i&=&C_{1i}x+C_{2i}, \nonumber \\
\phi_i&=&C_{1i}\textrm{cosh}({\eta _ix-4\eta
_i^3t})+C_{2i}\textrm{sinh}({\eta _ix-4\eta _i^3t}),\quad \eta
_i=\sqrt{-\lambda_i}\,,
\label{eq:generalphiiform1:pma-wcna2004}\nonumber \\
\phi_i&=&C_{1i}\sin (\eta _ix+4\eta _i^3t)+C_{2i}\cos (\eta
_ix+4\eta _i^3t),\quad \eta _i=\sqrt{\lambda_i}\,, \nonumber \eea
when $\lambda _i$ is zero, negative and positive, respectively.
Here $C_{1i}$ and $C_{2i}$ are arbitrary real constants.

More general Wronskian solutions can be constructed under a
broader set of sufficient conditions
\cite{SirianunpiboonHR-PLA1988}:
\[
-\phi_{i,xx}=\sum_{j=1}^i\lambda  _{ij}\phi _j,  \quad
\phi_{i,t}=-4\phi_{i,xxx},\quad 1\le i\le N,
\]
where the coefficient matrix $\Lambda :=(\lambda _{ij})$ is an
arbitrary constant lower-triangular matrix. Very recently, an
essential generalization to the above sufficient conditions is
presented in \cite{MaY-TAMS2004}:
\begin{equation}
 -\phi_{i,xx}=\D \sum_{j=1}^N\lambda  _{ij}\phi _j, \quad
\phi_{i,t}=-4\phi_{i,xxx}, \quad 1\le i\le N,
\end{equation}
 where the coefficient matrix $\Lambda =(\lambda _{ij})$ is an arbitrary real constant
 matrix, not being lower-triangular any more.
Once a set of eigenfunctions is obtained, the Wronskian solutions
to the KdV equation (\ref{eq:kdv:pma-wcna2004}) is given by
\begin{equation}
 u=-2\partial _x^2 \ln W(\phi_1,\phi_2,\cdots,\phi_N).
\end{equation}

It is easy to see that linear transformations of eigenfunctions do
not generate new Wronskian solutions. Therefore, we only need to
consider the Jordan form of the coefficient matrix $\Lambda$. A
real matrix can have and only have two types of Jordan blocks in
the real field:
\[\ba {rcl} \textrm{Type 1:} &\quad &\left[\ba {cccc}\lambda_i & & & 0
\vspace{2mm}\\
1 & \lambda_i  &  &  \vspace{2mm}\\
&\ddots& \ddots&  \vspace{2mm}\\
0 & & 1& \lambda_i  \ea \right]_{k_i\times k_i},  \vspace{2mm}
\\
 \textrm{Type 2:}&\quad &\left[\ba {cccc} A_i & & & 0
\vspace{2mm}\\
 I_2&A_i& & \vspace{2mm}\\
& \ddots& \ddots& \vspace{2mm}\\
0 &&I_2 &A_i \ea \right]_{l_i\times l_i},\quad A_i=\left[\ba
{cc}\alpha _i&-\beta _i\vspace{2mm}\\ \beta _i&\alpha _i \ea
\right],\ I_2=\left[\ba {cc}1&0\vspace{2mm}\\
0&1 \ea \right],  \ea \] where $\lambda_i$,  $\alpha_i $ and
$\beta_i
>0$ are real constants.
The first type of blocks only has the real eigenvalue $\lambda_i$
with algebraic multiplicity $k_i$, but the second type of blocks
has the complex eigenvalues $\lambda _{i}^{\pm} =\alpha_i\pm \beta
_i \sqrt{-1}$ with algebraic multiplicity $l_i$.

Note that an eigenvalue of the coefficient matrix $\Lambda
=(\lambda _{ij})$ is also an eigenvalue of the Schr\"odinger
operator $-\frac {\partial ^2}{\partial x^2}+u$ with zero
potential $u=0$. Therefore, according to the types of eigenvalues
of the coefficient matrix $\Lambda$, we can have four Wronskian
solution situations \cite{MaY-TAMS2004}:

\vskip 3mm
\begin{minipage}[t]{120mm}
\framebox[118mm]{\rule[-12mm]{0mm}{26mm} $ \ba {ll}
\textrm{Rational solutions:}&
\textrm{Type 1 with $\lambda _i=0$}, \vspace{1mm} \\
\textrm{Negaton solutions:}& \textrm{Type 1 with  $\lambda _i<0$},\vspace{1mm} \\
\textrm{Positon solutions:} & \textrm{Type 1 with  $\lambda _i>0$}, \vspace{1mm} \\
\textrm{Complexiton solutions:} & \textrm{Type 2 with complex
eigenvalues}.
 \ea  $}
\end{minipage}
\vskip 3mm

 {\bf Complexiton Solutions of Zero Order:}
 Assume that \begin{equation}
-\left[\ba {c}\phi _{i1,xx}  \vspace{2mm} \\
\phi_{i2,xx}\ea \right]= A_i
\left[\ba {c} \phi _{i1} \vspace{2mm}  \\
\phi_{i2}\ea \right], \ A_i=
\left[\ba {cc} \alpha _i &-\beta _i \vspace{2mm} \\
\beta _i& \alpha _i\ea \right],\quad
\left[\ba {c} \phi _{i1,t}  \vspace{2mm} \\
\phi_{i2,t}\ea \right]= -4\left[\ba {c}\phi _{i1,xxx}  \vspace{2mm} \\
\phi_{i2,xxx}\ea \right],\nonumber
 \end{equation}
where $\alpha_i$ and $\beta _i>0$ are arbitrary real constants.
Such eigenfunctions can be easily explicitly presented
\cite{Ma-PLA2002}, but it is not easy to get the eigenfunctions
associated with higher-order Jordan blocks of type 2 (see
\cite{MaY-TAMS2004} for more information).
 An
 $n$-complexiton solution of order $(l_1,\cdots,l_n)$ is defined
 by
\begin{equation}
  u=-2\partial _x^2 \ln W(
\phi_{11},\phi_{12},\cdots,\partial_{\alpha
_1}^{l_1}\phi_{11},\partial_{\alpha _1}^{l_1} \phi_{12}; \cdots;
\phi_{n1},\phi_{n2},\cdots,\partial_{\alpha
_n}^{l_n}\phi_{n1},\partial_{\alpha _n}^{l_n}\phi_{n2}),
\end{equation}
which corresponds to Jordan blocks of type 2. An $n$-complexiton
of order $(0,\cdots,0)$ ($n$-complexiton for short) is
\[ u=-2\partial _x ^2 \ln
W(\phi_{11},\phi_{12},\cdots,\phi_{n1},\phi_{n2}).
\]
In particular, a 1-complexiton \cite{Ma-PLA2002} is given by
 \bea u&=& -2\partial _x^2 \ln
W(\phi_{11},\phi_{12}) \nonumber
\\ &=& \D \frac {-4\beta _1^2\bigl[1+\cos(2\delta_1(x-\bar \beta
_1t)+2\kappa_{1})\cosh (2\Delta_1(x+\bar \alpha _1t)+2\gamma_{1})
\bigr]}{\bigl[\Delta _1 \sin (2\delta_1(x-\bar \beta
_1t)+2\kappa_{1})+\delta _1 \sinh (2\Delta_1(x+\bar \alpha
_1t)+2\gamma_{1})\bigr ]^2} \nonumber \\
&& + \D \frac {4 \alpha _1\beta _1\sin (2\delta_1(x-\bar \beta
_1t)+2\kappa_{1})\sinh (2\Delta_1(x+\bar \alpha _1t)+2\gamma_{1})
}{\bigl[\Delta _1 \sin (2\delta_1(x-\bar \beta
_1t)+2\kappa_{1})+\delta _1 \sinh (2\Delta_1(x+\bar \alpha
_1t)+2\gamma_{1})\bigr ]^2}, \nonumber \eea  where $\alpha
_1,\,\beta _1>0,\, \kappa_1$ and $\gamma_1$ are arbitrary
constants, and $\Delta _1,\, \delta _1,\, \bar {\alpha }_1,$ and
$\bar {\beta }_1$ are \bea && \Delta_1 =\sqrt{\frac {\sqrt{\alpha
_1^2+\beta _1^2}-\alpha_1 }{2}},\ \delta_1 =\sqrt{\frac
{\sqrt{\alpha _1^2+\beta _1^2}+\alpha _1}{2}}, \nonumber  \\ &&
\bar {\alpha }_1=4\sqrt{\alpha_1 ^2+\beta_1^2}+8\alpha _1 ,\ \bar
{\beta }_1=4\sqrt{\alpha_1^2+\beta_1^2}-8\alpha_1 . \nonumber \eea
The special case of $\alpha _1=0$ leads to the following solution
\[ u=\frac {8\beta _1+ 8\beta
_1\cos(\sqrt{2\beta_1}\,x-4\beta_1\sqrt{2\beta_1}\,t+2\kappa_{1})\cosh
(\sqrt{2\beta_1}\,x+ 4\beta_1\sqrt{2\beta_1}\, t+2\gamma_{1}) }
{\bigl[ \sin
(\sqrt{2\beta_1}\,x-4\beta_1\sqrt{2\beta_1}\,t+2\kappa_{1})+\sinh
(\sqrt{2\beta_1}\,x+4\beta_1\sqrt{2\beta_1}\,t+2\gamma_{1})\bigr
]^2}.\]
This solution is associated with purely imaginary eigenvalues of
the Schr\"odinger spectral problem with zero potential. Moreover,
the 1-complexiton above contains the breather-like or spike-like
solution presented in \cite{Jaworski-PLA1984}:
\[ u=\frac {8\{(a^2-b^2)(b/a)\cos \nu \sinh (\eta +p )+2b^2[1+\sin
\nu \cosh (\eta +p)]\}} {[\cos\nu -(b/a)\sinh (\eta +p)]^2},\]
where $\nu =-2bx+8(3a^2b-b^3)t,\, \eta =-2ax+8(a^3-3ab^2)t$ and $
p=\ln (b/a)$. This is obtained if we choose that \[ \kappa_1=\frac
{\pi}4,\ \gamma_1=\frac 12 \ln (\frac ba ),\ \Delta _1=-a ,\
\delta _1=b,\] where $a$ and $b$ are arbitrary real constants. Two
special cases of the 1-complexiton solution are depicted in Figure
1.~\begin{figure}[h] \centerline{
\epsfig{figure=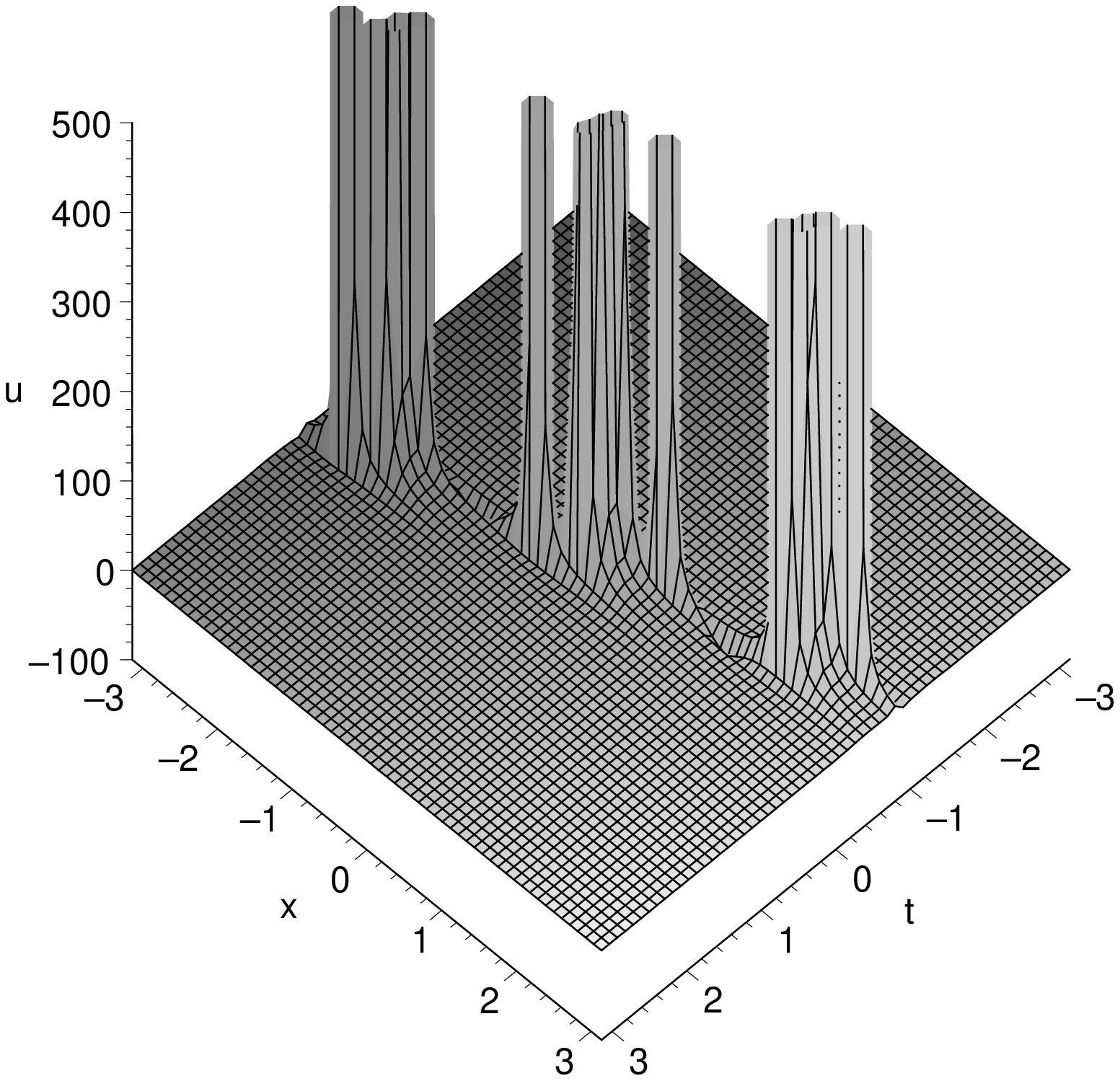, height=5.8cm, width=7.5cm}
\epsfig{figure=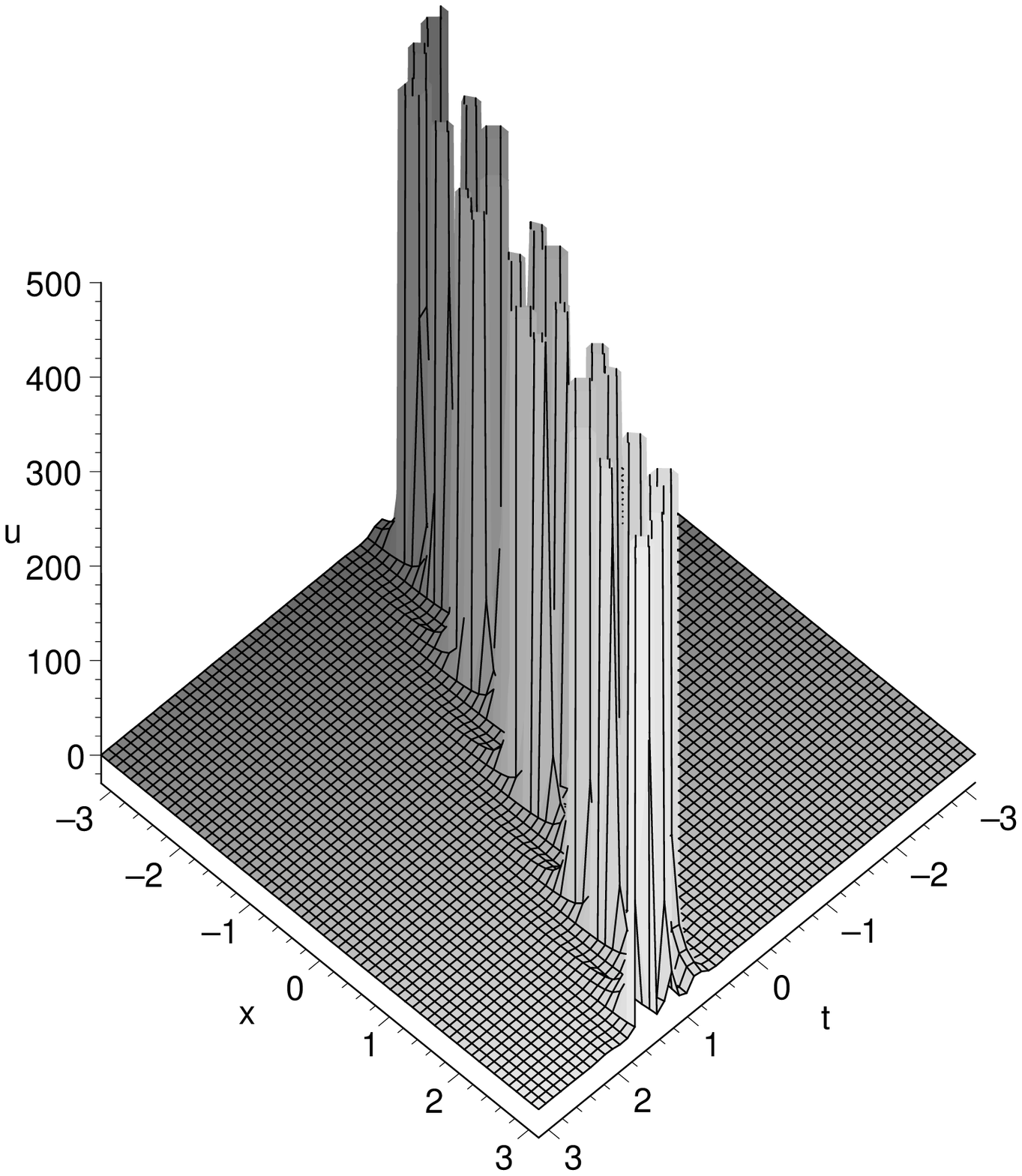, height=5.8cm, width=7.5cm}}
\vspace{-3mm} \caption{1-complexiton - $\alpha_1=0,\, \beta_1=1$
(left) $\quad $ 1-complexiton - $\alpha_1=1,\, \beta_1=1 $ (right)
} \label{fig:c1:rwcna04}
\end{figure}

Complexiton solutions of higher order are complicated but special
ones can be presented by taking derivatives of eigenfunctions with
respect to the involved parameters \cite{Ma-PLA2002,MaY-TAMS2004}.
Complexiton solutions are singular and they are not travelling
waves. It is clear that the 1-complexiton above is not a
travelling wave since $\beta _1\ne 0$. Moreover, based on the
Painlev\'e property of the KdV equation, the singularities of
complexitons are all poles of second order with respect to $x$.
This is obvious for the 1-complexiton above, since $u$ has no pole
singularity (the numerator of $u$ will be zero) if the function
$\Delta _1\sin \xi _1+\delta _1\sinh \xi_2$ is zero and its
spatial derivative $2\Delta_1\delta _2(\cos\xi _1 +\cosh \xi_2 )$
is also zero, where $\xi_1= 2\delta_1(x-\bar \beta
_1t)+2\kappa_{1}$ and $\xi_2=2\Delta_1(x+\bar \alpha
_1t)+2\gamma_{1}$.

\section{TODA LATTICE EQUATION}

Let us second consider the Toda lattice equation
\begin{equation}
\dot{a}_n=a_n(b_{n-1}-b_n) ,\quad \dot{b}_n=a_n-a_{n+1}
,\label{eq:Toda:pma-wcna2004}
\end{equation}
where $\dot{a}_n =\frac {d a_n}{dt}$ and $\dot{b}_n =\frac {d
b_n}{dt}$. Under
 the transformation
\begin{equation}
a_n=1+\frac{d^2}{dt^2}\log
\tau_n=\frac{\tau_{n+1}\tau_{n-1}}{\tau_n^2}, \quad
b_n=\frac{d}{dt}\log \frac{\tau_{n}}{\tau_{n+1}}=\frac {\dot
{\tau}_{n}\tau_{n+1}-\tau_n\dot{\tau}_{n+1} }{\tau_n{\tau}_{n+1}},
 \label{eq:BTToda:pma-wcna2004} \end{equation}
the Toda lattice equation (\ref{eq:Toda:pma-wcna2004}) becomes
\[
\bigl[D_t^2-4\sinh ^2(\frac {D_n}{2})\bigr ]\tau_n\cdot \tau_n=0,
\]
 where $D_t$ and $D_n$ are Hirota's operators.
That is, \begin{equation}
 \ddot{\tau}_n\tau
 _n-(\dot{\tau}_n)^2-\tau_{n+1}\tau_{n-1}+\tau_n^2=0.\label{eq:bToda:pma-wcna2004}
  \end{equation}

The Casoratian technique \cite{Nimmo-PLA1983} is one of the ways
to solve the bilinear Toda lattice equation
(\ref{eq:bToda:pma-wcna2004}). The corresponding solutions to the
bilinear Toda lattice equation (\ref{eq:bToda:pma-wcna2004}) are
determined by the Casorati determinant:
\begin{equation}
 \tau_n=\textrm{Cas}(\phi _1,\phi_2,\cdots,\phi_{N})
 :=\left |
\ba {cccc}\phi _1(n) &\phi _1{(n+1)}&\cdots &\phi_1{(n+N-1)}\\
\phi _2(n) &\phi _2{(n+1)}&\cdots &\phi_2{(n+N-1)}\\
  \vdots &\vdots & \ddots & \vdots \\
  \phi _N(n) &\phi _N{(n+1)}&\cdots &\phi_N{(n+N-1)}
  \ea \right |,
  \end{equation}
and the Casoratian technique requires \cite{MaM-PA2004}
\begin{equation}
  \phi_i(n+1)+\phi_i(n-1)=\D \sum_{j=1}^N \lambda _{ij} \phi_j(n),
\quad
\partial_t \phi_{i}(n)=\phi_i(n+\delta )
,\quad 1\le i\le N,
\end{equation}
where $\delta =\pm 1 $ and $\lambda_{ij}$ are arbitrary real
constants. Similarly, we only need to consider the Jordan form for
the coefficient matrix $\Lambda =(\lambda_{ij})$. Assume that two
types of Jordan blocks of the coefficient matrix $\Lambda
=(\lambda _{ij})$ are specified as in the case of the KdV equation
in the last section. Then, based on the solution structures of
eigenfunctions associated with eigenvalues of different types, we
can similarly have four Casoratian solution situations:

\vskip 3mm
\begin{minipage}[t]{120mm}
\framebox[118mm]{\rule[-12mm]{0mm}{26mm} $ \ba {ll}
\textrm{Rational solutions:}&
\textrm{Type 1 with $\lambda _i=\pm 2$},  \vspace{1mm} \\
\textrm{Negaton solutions:}& \textrm{Type 1 with  $\lambda _i<-2$}\ \textrm{or}\ \lambda_i>2, \vspace{1mm} \\
\textrm{Positon solutions:} & \textrm{Type 1 with  $-2<\lambda _i<2$}, \vspace{1mm} \\
\textrm{Complexiton solutions:} & \textrm{Type 2 with complex
eigenvalues.}
 \ea  $}
\end{minipage}

\vskip 3mm

 {\bf Case $k_i=1$ of Type 1}: We have
 \begin{equation} \phi_i(n+1)+\phi_i(n-1)=\lambda _i \phi_i(n) ,\quad
\part _t \phi_{i}(n)=\phi_i(n+\delta ), \label{eq:Kiis1Type1Toda:pma-wcna2004} \end{equation}
 where $ \delta =\pm 1$ and $\lambda _i=\textrm{consts.}$
Its general
 eigenfunctions are: \bea
&&\phi_i=C_{1i} (\delta n+ \varepsilon t) \varepsilon ^n
\textrm{e}^{\varepsilon t}+ C_{2i}\varepsilon ^n
\textrm{e}^{\varepsilon t}
,\quad  \lambda _i=2 \varepsilon, \nonumber \\
 &&\phi_i=C_{1i}\textrm{e}^{t\cos k_i}\cos
({k_in+\delta t\sin k_i }) +C_{2i} \textrm{e}^{t\cos k_i}\sin
({k_in+\delta t\sin k_i }) , \quad \lambda _i=2\cos k_i
\nonumber \\
&& \phi_i=C_{1i} \varepsilon ^n \textrm{e}^{k_in+\varepsilon
t\textrm{e}^{\delta k_i }} +C_{2i}\varepsilon ^n \textrm{e}^{-k_in
+\varepsilon t \textrm{e}^{-\delta k_i}} , \quad
 \lambda_i=2\varepsilon \cosh k_i ,\nonumber
 \eea
where $\varepsilon =\pm 1$, $|\cos k_i|<1$ for the second type of
eigenfunctions, $\cosh k_i>1$ for the third type of
eigenfunctions, and $C_{1i}$ and $C_{2i}$ are arbitrary real
constants. These three types of eigenfunctions lead to rational
solutions (see, for example, \cite{MaY-CSF2004}), negatons and
positons (see, for example, \cite{MarunoMO-JPSJ2004}),
respectively.

{\bf Case $l_i=1$ of Type 2}: To construct complexitons, we solve
\begin{equation} \left \{ \ba {l} \phi_1(n+1)+\phi_1(n-1)=\alpha
\phi_1(n)-\beta \phi_2(n),
 \vspace{2mm}\\
\phi_2(n+1)+\phi_2(n-1)=\beta \phi_1(n)+\alpha \phi_2(n),
 \ea \right. \quad
\left\{\ba {l} \part _t \phi_1(n)=\phi_1(n+\delta ),\vspace{2mm}\\
\part _t \phi_2(n)=\phi_2(n+\delta ),\ea \right.
\label{eq:representativesysytemforcomplexitonsolToda:pma-wcna2004}
\end{equation} where $\delta =\pm 1$ and $\alpha, \beta=\textrm{consts.}$
Assume that
\[\alpha +\beta \sqrt{-1}\,=\lambda =2\cosh k =\textrm{e}^k+\textrm{e}^{-k},\ k=a+b\sqrt{-1}\,,\]
where $a$ and $b$ are real, and then
\[\alpha =2
\cosh a\cos b,\ \beta = 2 \sinh a\sin b .\] View
(\ref{eq:representativesysytemforcomplexitonsolToda:pma-wcna2004})
as a compact equation like (\ref{eq:Kiis1Type1Toda:pma-wcna2004})
where $\lambda =a+b\sqrt{-1}$, and then using the third type of
eigenfunctions above, we can generate the corresponding
eigenfunctions of
(\ref{eq:representativesysytemforcomplexitonsolToda:pma-wcna2004}):
\begin{equation}
 \left\{ \ba {l} \phi_1(n) = c\textrm{e}^{ na+
t\textrm{e}^{\delta a}\cos \delta b}\cos( n b+ t\textrm{e}^{\delta
a}\sin \delta b )   +d\textrm{e}^{- n a+ t\textrm{e}^{-\delta
a}\cos \delta b }\cos( n b+ t\textrm{e}^{-\delta a}\sin \delta b
),
\vspace {2mm}\\
\phi_2(n) = c\textrm{e}^{ n a+ \textrm{e}^{\delta a}\cos \delta
b}\sin( n b+ t\textrm{e}^{\delta a}\sin \delta b ) -d\textrm{e}^{-
n a+ t\textrm{e}^{-\delta a}\cos \delta b }\sin( n
b+t\textrm{e}^{-\delta a}\sin \delta b ).
 \ea \right.
 \label{eq:formulasforphi_1andphi_2:pma-wcna2004}\end{equation}
This set of eigenfunctions yields the $\tau$-function of
1-complexiton:
\begin{eqnarray}
 \tau_n &=&\textrm{Cas}(\phi_1(n),\phi_2(n))\nonumber \\
 & =&
2cd \,\textrm{e}^{2t\cosh \delta a \cos \delta b
}\sin(2nb+ b+2t\cosh \delta a \sin \delta b  )\sinh a \nonumber \\
&& + c^2\textrm{e}^{2na+a+2t\textrm{e}^{\delta a}\cos \delta b
}\sin b-d^2\textrm{e}^{-2na-a+2t\textrm{e}^{-\delta a}\cos
 \delta b }\sin b\,,
 \label{eq:taufunctionforsinglecomplexiton:pma-wcna2004}\end{eqnarray}
where $\delta=\pm 1$ and $a,b,c,d=\textrm{consts.}$
 In particular, $c=\pm d$ leads to the $\tau$-function:
\begin{eqnarray}
 \tau_n &=&2c^2 \, \textrm{e}^{2t\cosh \delta a \cos \delta b }\sinh
(2na+a+2t\sinh \delta a \cos \delta b)\sin b \nonumber  \\ && \pm
2c^2 \,\textrm{e}^{2t\cosh \delta a \cos \delta b }\sin(2nb+
b+2t\cosh \delta a \sin \delta b  )\sinh a \,.\nonumber
\label{eq:simplifiedtaufunctionforsinglecomplexiton:pma-wcna2004}\end{eqnarray}
The resulting 1-complexiton solution of the Toda lattice equation
 (\ref{eq:Toda:pma-wcna2004}) is given by
 (\ref{eq:BTToda:pma-wcna2004})
where $\tau _n$ is presented in
(\ref{eq:taufunctionforsinglecomplexiton:pma-wcna2004}). Two
special cases with $\delta=c=d=1$ of this 1-complexiton solution
are depicted in Figures 2 and 3.~\begin{figure}[h] \centerline{
\epsfig{figure=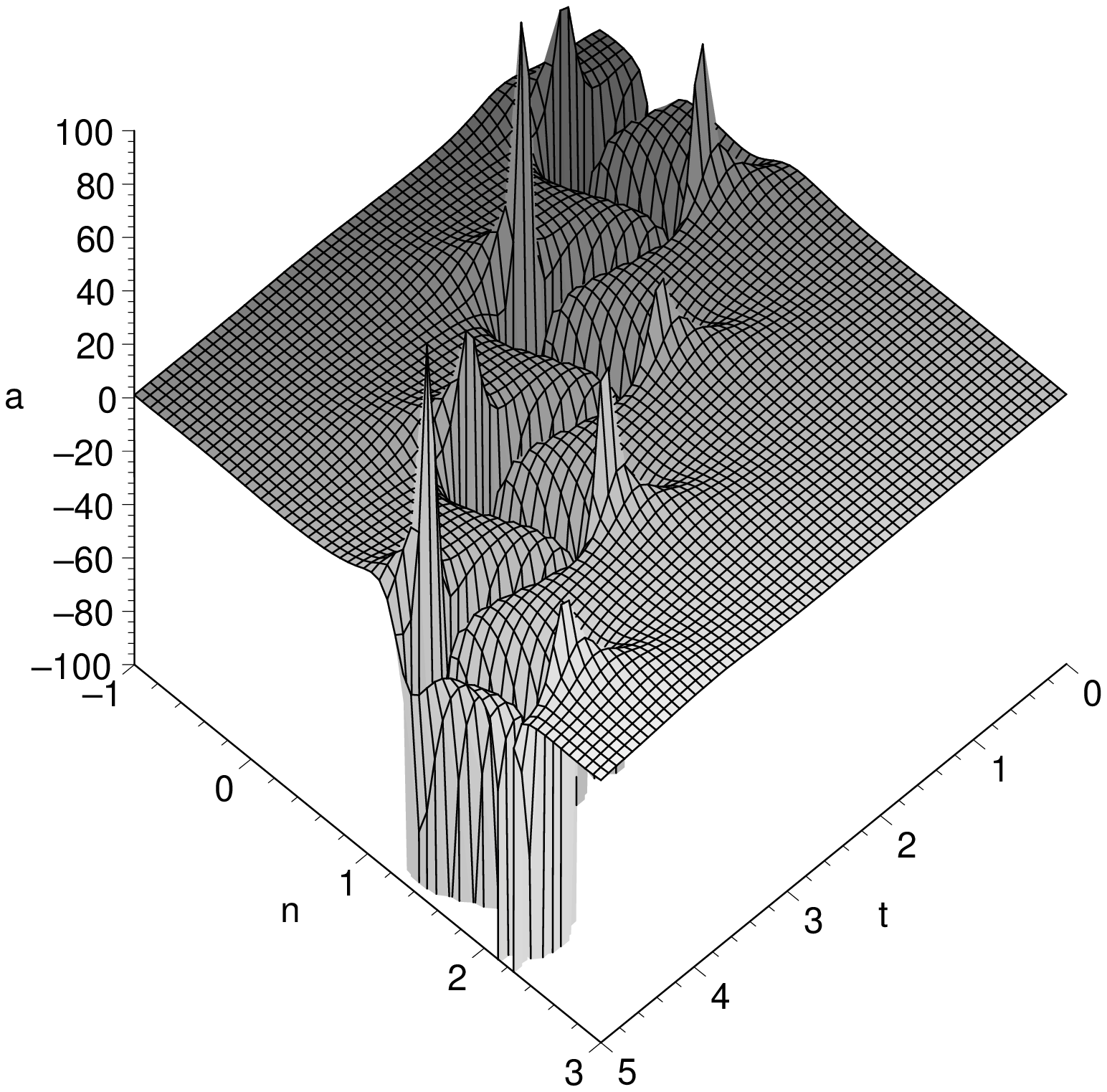, height=5.9cm, width=7.5cm}
\epsfig{figure=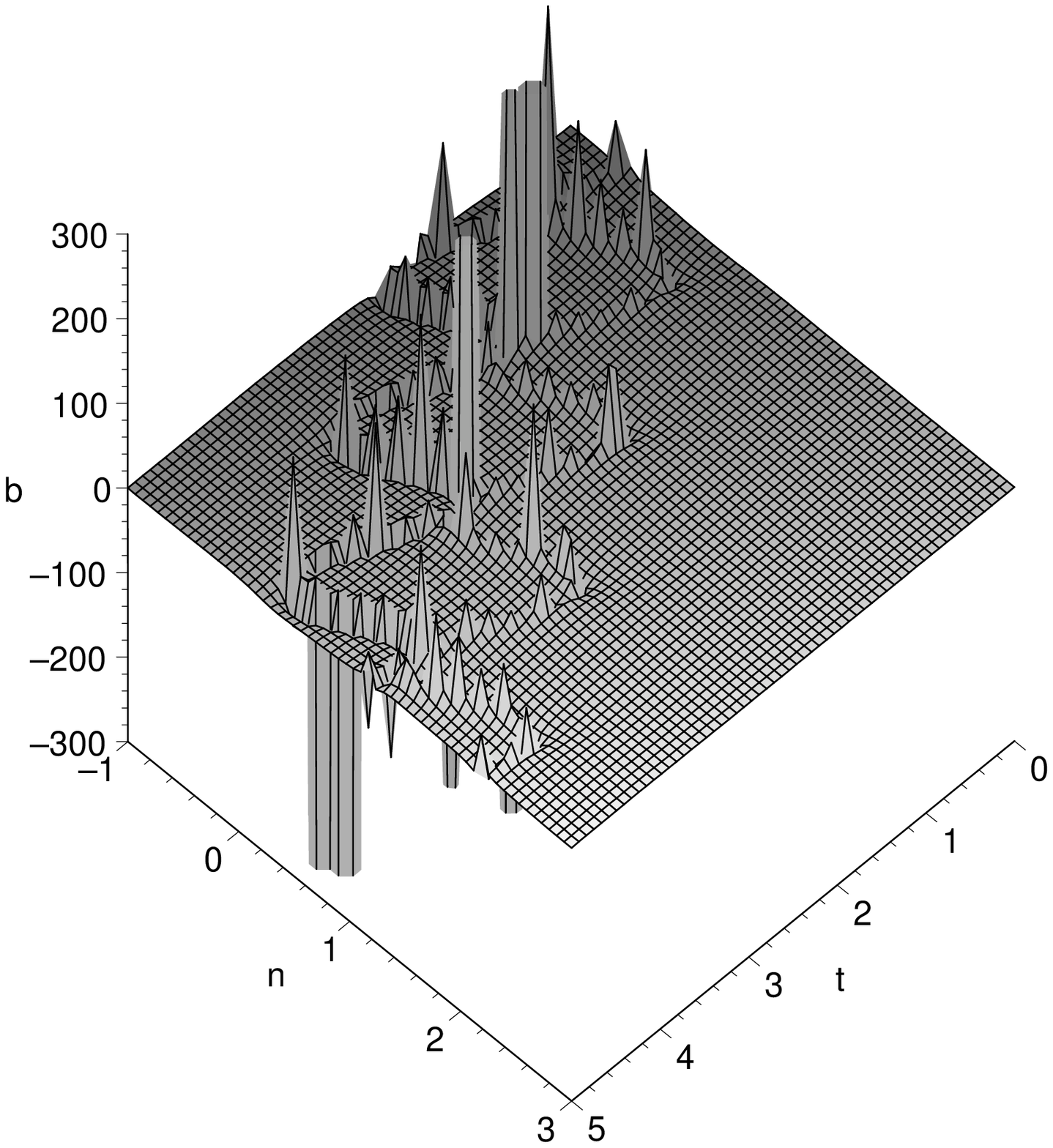, height=5.9cm, width=7.5cm}}
\vspace{-3mm}
 \caption{1-complexiton $a_n$ - $2a=b=2$ (left)
$\quad $1-complexiton $b_n$ - $2a=b=2$ (right) }
\label{fig:1stcaseTodacomplexiton:pma-wcna04}
\end{figure}~\begin{figure}[h] \centerline{ \epsfig{figure=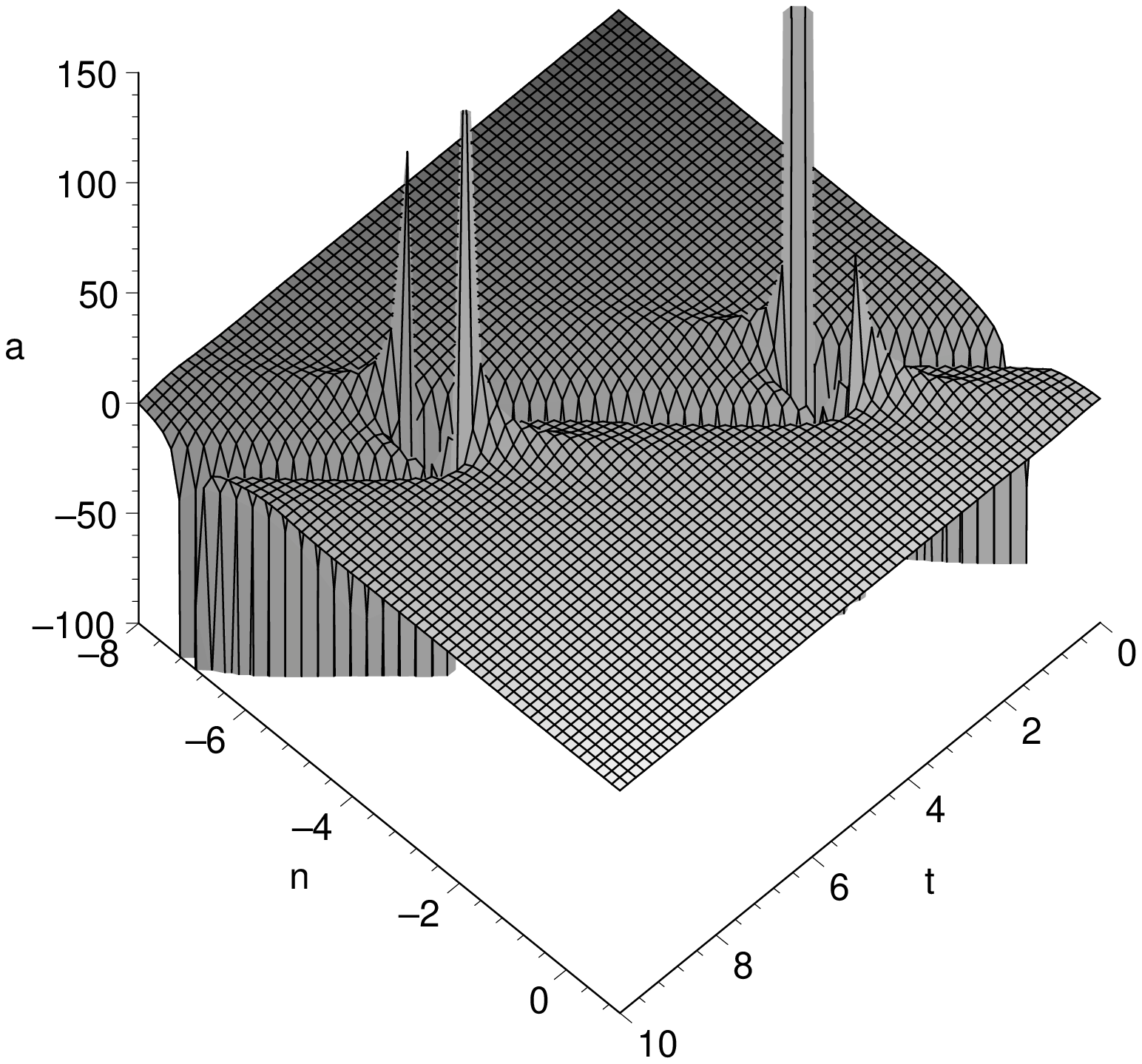,
height=5.9cm, width=7.5cm} \epsfig{figure=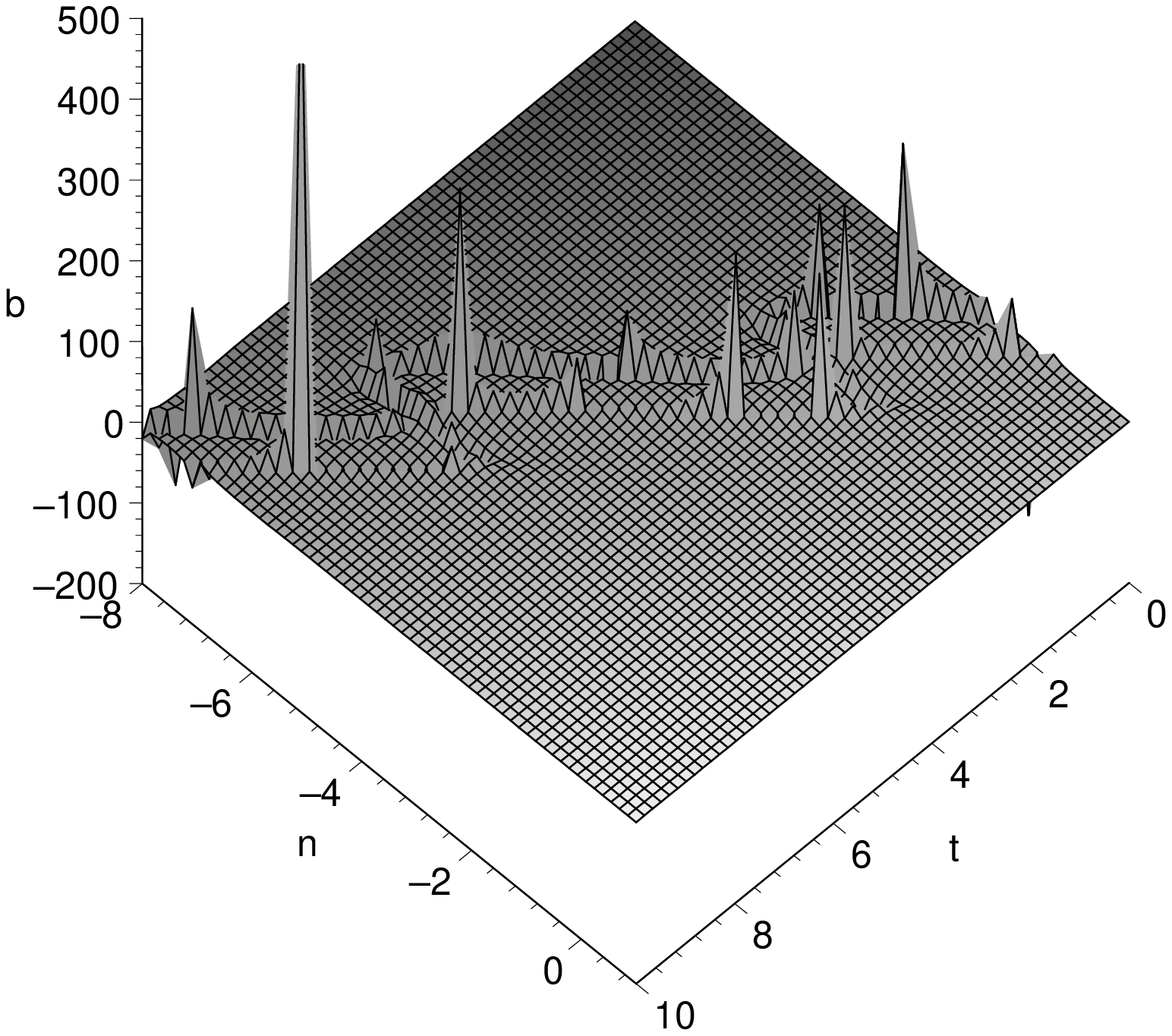,
height=5.9cm, width=7.5cm}} \vspace{-3mm}
 \caption{1-complexiton
$a_n$ - $a=-b=1$ (left) $\quad $1-complexiton $b_n$ - $a=-b=1$
(right) } \label{fig:2ndcaseTodacomplexiton:pma-wcna04}
\end{figure}
Complexiton solutions of higher order are complicated but special
ones can be constructed by computing derivatives of eigenfunctions
with respect to the involved parameters \cite{MaM-PA2004}.

\section{CONCLUDING REMARKS}

There is a correspondence between the characteristic linear
problems of the KdV equation and the Toda lattice equation:
\[\ba {rcl} \partial _x^2
  & \Leftrightarrow &
 \phi(n+1)+\phi(n-1) -2\phi(n),
\vspace{2mm}\\
-\partial _x^2\phi=\lambda \phi   & \Leftrightarrow &
 \phi(n+1)+\phi(n-1) =\lambda \phi(n),
\vspace{2mm}\\
-\partial _x^2\phi=\lambda \phi   & \Leftrightarrow &
 -\phi(n+1)-\phi(n-1)+2\phi(n) =(-\lambda+2) \phi(n),
\ea \] and a correspondence between the eigenvalues of the
characteristic linear problems:
\[\ba {lcr}
 \lambda =0,>0,<0\ \textrm{and $\lambda =$ complex}
& \Leftrightarrow  & |\lambda| =2,<2,>2\ \textrm{and $\lambda =$
complex},\ \textrm{respectively}. \ea
\]
We can also observe the correspondence between the KdV equation
and its characteristic linear problem $ -\partial _x ^2 \phi
=\lambda \phi $. The characteristic equation of $ -\partial _x ^2
\phi =\lambda \phi $ is
\[ -m^2-\lambda =0 .\]
Therefore, for example, we see that
\[\ba {rcl} \lambda \
\textrm{is positive}
  &\Leftrightarrow  & m \ \textrm{is purely
imaginary},\vspace{2mm}\\
 \lambda \ \textrm{is complex}
& \Leftrightarrow &   m \ \textrm{is complex but not purely
imaginary}. \ea
\]
Together with other obvious correspondences, this implies that
\[
\ba {rcl} \lambda =0   & \Leftrightarrow  &   \textrm{KdV rational
solutions}, \vspace{2mm}
\\
\lambda <0   &  \Leftrightarrow &    \textrm{KdV solitons and
negatons},
\vspace{2mm}\\
 \lambda >0   &  \Leftrightarrow  &   \textrm{KdV
positons}, \vspace{2mm}\\
\lambda= \ \textrm{complex}   & \Leftrightarrow &   \textrm{KdV
complexitons.} \ea
\]
 It then follows that the obtained complexiton solutions to the KdV
equation and the Toda lattice equation satisfy the two criteria
for complexiton solutions stated in the introduction, indeed.
Because of the characteristic in the second criterion,
complexitons may play a role similar to the one that the imaginary
unit $\sqrt{-1}$ plays in physics.

We also mention that the Wronskian determinants and the Casorati
determinants give many other solutions if different types of
eigenvalues are allowed. Such solutions are called interaction
solutions among determinant solutions of different kinds.
 For higher dimensional integrable equations, the solution
situations are much more diverse
\cite{KasmanG-JMP2001,TangCL-JPA2002} and the problem of
classifying solutions is extremely difficult \cite{Ma-PLA2003}. It
is hoped that the study of complexitons could further assist in
understanding, identifying and classifying nonlinear integrable
differential equations and their exact solutions.

\section*{Acknowledgment}
The author is grateful to Profs. M. Gekhtman, M. Kovalyov, S.Y.
Lou and K. Maruno for enthusiastic comments and valuable
discussions during WCNA2004.

\end{document}